\documentclass[10pt,pra,twocolumn,superscriptaddress,amsfonts,amssymb,amsmath,aps,floatfix]{revtex4-2}
\usepackage{graphicx}
\usepackage{braket}
\usepackage{amsmath}
\usepackage{color}
\usepackage{orcidlink}
\usepackage{appendix}
\usepackage{dsfont}

\usepackage{hyperref}

\def\adag{\hat{a}^{\dag}}
\def\a{\hat{a}}
\def\sz{\hat{\sigma}_z}
\def\sx{\hat{\sigma}_x}

\def\tr{\operatorname{tr}}
\def\erf{\text{erf}}

\begin{document}

\title{Convergence to semiclassicality in the quantum Rabi model}

\author{H.~F.~A.~Coleman\,\orcidlink{0009-0003-2607-4382}}
\thanks{Contact author: \href{mailto:hfac1g20@soton.ac.uk}{hfac1g20@soton.ac.uk}}
\affiliation{School of Physics and Astronomy, University of Southampton,
Highfield, Southampton, SO17 1BJ, United Kingdom}

\author{R.~A.~Morrison\,\orcidlink{0009-0001-3574-1146}}
\altaffiliation{Present address: Department of Physics and Astronomy,
University College London, London WC1E 6BT, United Kingdom.}
\affiliation{School of Physics and Astronomy, University of Southampton,
Highfield, Southampton, SO17 1BJ, United Kingdom}

\author{A.~D.~Armour\,\orcidlink{0000-0002-7190-8601}}
\affiliation{School of Physics and Astronomy and Centre for the Mathematics and
Theoretical Physics of Quantum Non-Equilibrium Systems, University of Nottingham,
Nottingham NG7 2RD, United Kingdom}

\author{E.~K.~Twyeffort\,\orcidlink{0000-0003-0113-0977}}
\affiliation{School of Physics and Astronomy, University of Southampton,
Highfield, Southampton, SO17 1BJ, United Kingdom}

\date{\today}

\begin{abstract}
We investigate the emergence of semiclassical dynamics in the quantum Rabi model using a recently developed limiting procedure that formally establishes correspondence with the semiclassical Rabi Hamiltonian [E. K. Twyeffort Irish and A. D. Armour, Phys. Rev. Lett. 129, 183603 (2022)]. While the limit itself is defined at the Hamiltonian level, how it is reached depends on the choice of quantum states. Defining a set of quantitative measures that capture the differences between quantum and semiclassical dynamics, we examine convergence to the semiclassical limit when the field is prepared in a displaced number state. These states, which interpolate to Fock states for zero displacement, are more general than the set of coherent states usually employed when considering the emergence of semiclassical behavior. Numerical computations of these measures consistently demonstrate the progressive emergence of semiclassical behavior as the joint limit of vanishing coupling and infinite displacement is approached. Complementing the numerical results, analytical approximations are developed that reproduce the behavior in the vicinity of the semiclassical limit with a high degree of fidelity and allow scaling relations to be derived. Although any initial displaced number state will eventually converge to the corresponding semiclassical dynamics as the limit is taken, the rate of convergence depends on the Fock number $n$ of the state. States with larger values of $n$, which behave less classically than coherent states, converge more slowly to the limit.
\end{abstract}

\maketitle
\section{Introduction}\label{sec:intro}

A bipartite quantum system is said to behave semiclassically when the properties of one quantum subsystem can be predicted accurately using a simpler semiclassical model in which the other subsystem is treated classically. Determining exactly when semiclassical behavior emerges is a significant theoretical challenge, even in apparently very simple systems, and is closely related to the more general problem of quantum-classical correspondence. The most universal approach for reconciling quantum and classical predictions is taking the limit $\hbar \to 0$~\cite{FeynmanHibbs1965, Berry1984}. However, this limit does not suffice in the semiclassical case where one subsystem remains quantum. Relaxing the fully quantum dynamics to a simpler semiclassical approximation is common in quantum optics~\cite{Bishop2010, Musa2019, Kurko2024, Jurgens2021} and similar simplifications are also applied in quantum chaos~\cite{Emary2003, Muller2004}, quantum thermodynamics~\cite{Giorgini1997} and condensed matter physics~\cite{Salkola1995, Lee1994, Lill1989}. Joint scaling limits address the problematic $\hbar \to 0$ limit by scaling multiple system parameters together, with the scaling chosen to yield a well-defined, nontrivial limit --- for example, by keeping a chosen product or quotient finite. This technique allows a single subsystem to admit a classical description while the other remains fully quantum~\cite{Layton2024, Correggi2024, Colla2024, cavina2024}.




The quantum Rabi model (QRM), which describes the interaction between a spin-$1/2$ particle (spin) and a single electromagnetic field mode~\cite{Rabi1937}, is a natural setting for exploring the emergence of semiclassical dynamics. It is one of the simplest models in which the same fundamental interaction can be described either with a fully quantized field or with an effective classical drive. In recent years, rapid progress in engineered solid-state devices has renewed interest in this problem~\cite{Fink2010, Vlastakis2013}, providing direct access to experiments where both quantum and semiclassical regimes are accessible. While many cavity and circuit quantum electrodynamics (QED) experiments require a fully quantum-mechanical description of the field~\cite{Haroche1996, Wineland1996, Blais2004}, other experimental architectures routinely operate in regimes where the field is assumed to behave as a classical drive~\cite{Golter2014, Leibfried2003, Stievater2001}. 

A recent Letter by two of the present authors \cite{Irish2022} links the quantum and semiclassical Rabi models by showing that the semiclassical version can be obtained directly from its quantum counterpart in an appropriate joint scaling limit. This mathematical formalism operates at the Hamiltonian level, without explicitly imposing specific conditions on the initial state of the field. Intriguingly, however, it implies that a coherent state of the field is not required to recover semiclassical dynamics; rather, \emph{any} displaced Fock state will give rise to the same semiclassical dynamics under the joint limit. 

In this work, we test the predictions of \cite{Irish2022} by exploring the dynamics of the Rabi model with different displaced Fock states of the field. Three criteria for genuine semiclassical behavior are identified: (i) the spin dynamics reduces to that generated by the corresponding semiclassical Hamiltonian; (ii) no entanglement is generated between the spin and field; and (iii) the state of the field is unaffected by the spin and evolves classically, in the sense that the expectation values of its quadrature operators follow the classical equations of motion. We determine numerically how each of these properties changes as the joint limit is approached. Analytical approximations are also derived, providing insight into the physical mechanisms underpinning the quantum-to-semiclassical transition. Our results establish that the emergence of semiclassical behavior does not depend on the degree of classicality of the field, but the rate of convergence to the limit does.

We begin in Sec.~\ref{sec:sc-limit} by connecting the Hamiltonian-level formalism laid out in~\cite{Irish2022} with the properties of individual initial states of the field. The features that define the semiclassical regime are identified and discussed. In Sec.~\ref{sec:semiclassical-measures} we present a detailed numerical and analytical study of the convergence to the semiclassical limit. Metrics are constructed for each of the identified aspects of semiclassical behavior. Using a numerical solution of the full Hamiltonian, the dependence of each metric on the coupling $\lambda$ is computed with different initial states of the field. The numerical results are complemented by analytical expressions obtained from quantum-corrected Floquet dynamics (QCFD) in the Floquet-basis rotating-wave approximation (FBRWA)~\cite{TwyeffortArmour2025}. Scaling relationships are extracted, providing support for the assertion that the rate of convergence to the semiclassical limit for an initial displaced Fock state $\ket{\alpha,n}$ of the field varies as $1/\sqrt{n}$. Finally, in Sec.~\ref{sec:discussion} we conclude.

\section{Emergence of semiclassical behavior}\label{sec:sc-limit}

The quantum Rabi Hamiltonian is defined as 
\begin{equation}
    \hat{H}_q = \frac{\Omega}{2} \sz + \omega_0 \adag \a + \lambda (\adag + \a) \sx ,
\label{eq:Hq}
\end{equation}
where $\Omega$ is the energy splitting of the spin, $\omega_0$ is the frequency of the field, and $\lambda$ is the single-photon quantum coupling. The field operators $\adag$ and $\a$ are the creation and annihilation operators, respectively, and $\hat{\sigma}_{z, x}$ are the standard Pauli spin operators; we have also set $\hbar =1$. The simpler Jaynes-Cummings model (JCM) is obtained from Eq.\ \eqref{eq:Hq} by making a rotating wave approximation (RWA). This involves dropping the counter-rotating terms from the spin-field interaction, $\a\hat{\sigma}_-$ and $\adag\hat{\sigma}_+$, where $\hat{\sigma}_-$ ($\hat{\sigma}_+$) is the spin lowering (raising) operator\,\cite{LarsonMavrogordatos}. 

The semiclassical limiting procedure outlined in \cite{Irish2022} begins by transforming the Hamiltonian to an interaction picture with respect to the field. Another unitary transformation, the displacement $\hat{D}(\alpha) = \exp [\alpha \adag - \alpha^* \a]$, is then applied to the interaction picture Hamiltonian. Following the transformations, the Hamiltonian takes the form 
\begin{equation}
\begin{split}
        \hat{H}_q^I(t) = \frac{\Omega}{2} \sz &+ \lambda (e^{i \omega_0 t} \alpha^* + e^{-i \omega_0 t} \alpha) \sx \\
        &  + \lambda (e^{i \omega_0 t} \adag + e^{-i \omega_0 t} \a) \sx .
    \label{eq:transformed-Hq}
\end{split}
\end{equation}
The final step of the procedure requires taking the joint limit $\lambda \to 0$ and $|\alpha| \to \infty$ while keeping their product fixed. This removes the term involving field operators in Eq.~\eqref{eq:transformed-Hq}. Identifying the product $\lambda|\alpha|$ with the semiclassical drive amplitude $A$, the result of carrying out the limit is a Hamiltonian of the form
\begin{equation}
    \lim_{|\alpha| \to \infty, \lambda \to 0} \hat{H}_q^I(t) = \hat{H}_{sc}(t) \otimes \hat{I}_f,
\label{eq:Hq-limit}
\end{equation}
where $\hat{H}_{sc}(t)$ is the semiclassical Rabi Hamiltonian and $\hat{I}_f$ denotes the identity operator for the field. 


This formulation of the semiclassical limit is carried out at the Hamiltonian level and without imposing any explicit assumptions regarding the state of the field. To draw a connection between this approach and the properties of individual states, we first note that applying the displacement transformation $\hat{D}(\alpha)$ to the Hamiltonian is mathematically equivalent to expressing the Hamiltonian in the basis of displaced Fock states $\ket{\alpha, n} = \hat{D}(\alpha)\ket{n}$, where the Fock states $\ket{n}$ are the eigenstates of the field Hamiltonian $\omega_0 \hat{a}^{\dag} \hat{a}$. The expectation values of the interaction-picture quadrature operators
\begin{align}
    \hat{q}(t) &= \sqrt{\frac{\hbar}{2\omega_0}} (\adag e^{i \omega_0 t} + \a e^{-i \omega_0 t}) , \\
    \hat{p}(t) &= i\sqrt{\frac{\hbar \omega_0}{2}} (\adag e^{i \omega_0 t} - \a e^{-i \omega_0 t})     
\end{align}
are identical for all $\ket{\alpha,n}$ with the same $\alpha$:
\begin{align}
    \bra{\alpha,n} \hat{q}(t) \ket{\alpha,n} &= \sqrt{\frac{\hbar}{2\omega_0}} (\alpha^* e^{i \omega_0 t} + \alpha e^{-i \omega_0 t}) , \\
    \bra{\alpha,n} \hat{p}(t) \ket{\alpha,n} &= i \sqrt{\frac{\hbar \omega_0}{2}} (\alpha^* e^{i \omega_0 t} - \alpha e^{-i \omega_0 t}) .
\end{align}
In accordance with Ehrenfest's theorem, the expectation values of these observables follow the classical equations of motion. This allows the displacement parameter to be interpreted as
\begin{equation}
    \alpha = \frac{1}{2} \frac{\langle q(0)\rangle}{\Delta q_0} + \frac{i}{2} \frac{\langle p(0)\rangle}{\Delta p_0} ,
\end{equation}
where $\Delta q_0 = \sqrt{\hbar/(2\omega_0)}$ and $\Delta p_0 = \sqrt{\hbar \omega_0/2}$ are the ground-state uncertainties. Hence the limit $\lvert \alpha \rvert \to \infty$ implies that the quadrature expectation values --- which determine the field's displacement in phase space --- become macroscopically large relative to the ground-state uncertainties. This represents a classical limit for the field~\footnote{Further subtleties of this interpretation will be explored in upcoming work.}.

Remarkably, this argument implies that every displaced Fock state $\ket{\alpha, n}$ with the same $\alpha$ induces identical spin dynamics. This runs counter to the widespread assumption that the degree of classicality of a field is determined by its average photon number $\langle n \rangle$. What is more, it suggests that recovering semiclassical behavior of the spin does not require the field to be in its most classical quantum state, a coherent state. Nonclassical field states --- as measured by, e.g., negativity of the Wigner function --- can still give rise to semiclassical dynamics. However, the analysis sketched out in \cite{Irish2022} indicates that field states $\ket{\alpha,n}$ with higher values of $n$ will be more strongly affected by quantum field corrections to the semiclassical dynamics and further predicts that the contribution of the correction terms should scale as $\lambda \sqrt{n}$ to leading order. Hence we expect that the rate of convergence to the semiclassical limit for a given initial field state $\ket{\alpha,n}$ depends on $n$. 

In order to explore this hypothesis, we must first characterize the properties of semiclassical behavior. Comparisons between quantum and semiclassical dynamics in quantum optics have traditionally focused on the population dynamics of the spin. However, this provides an incomplete picture: for example, if the field is initially prepared in a Fock state, the spin population in the JCM exhibits sinusoidal Rabi oscillations resembling the semiclassical prediction\,\cite{LarsonMavrogordatos}, despite the fact that the underlying dynamics are fully quantum and the field possesses no classical analog. To complete the picture, we must take into account the behavior of both the spin and the field. The full quantum state of the spin --- not just its populations --- should evolve according to the corresponding semiclassical model. Since the effect of the field on the spin reduces to a classical driving term in the semiclassical limit, the spin and field must obviously remain unentangled throughout the evolution. 

But what of the field itself? Equation~\eqref{eq:Hq-limit} indicates that in the semiclassical limit, the quantum state of the field is unaffected by the coupling to the spin. As we are in the interaction picture with respect to the field, any initial field state will evolve according to its bare Hamiltonian, with the quadrature expectation values following the classical equations of motion. There is no need to specify a particular form for the field state within this formulation of the semiclassical limit. Hence our definition of `classicality' for the field differs from the various measures used in quantum optics that are based on how closely a given state resembles a coherent state, such as negativity of the Wigner function and the Mandel $Q$-parameter~\cite{Mandel_Wolf_1995}.

\section{Convergence to semiclassicality of quantum states}\label{sec:semiclassical-measures}

To explore the convergence to the semiclassical limit, we construct metrics for each of these characteristics of semiclassical behavior and examine their dependence on $\lambda$ and $\lvert \alpha \rvert$. The trace distance is used to compare the states of the spin and of the field at a given time with the predictions of the semiclassical model. A complementary measure that captures the similarity between the quantum and semiclassical population dynamics of the spin over an extended time period is provided by the Pearson correlation coefficient between the atomic inversion spectra. Entanglement is measured by the average von Neumann entropy over a given time period.

An appreciation of the relevant timescales in the problem helps to frame the dynamics. In the resonant JCM with the field in a coherent state, the spin dynamics exhibits three distinct timescales~\cite{Eberly1980,LarsonMavrogordatos}. The collapse time scales as $\tau_\text{col}\propto 1/\lambda$, while the revival time scales as $\tau_\text{rev}\propto |\alpha|/\lambda$. Both of these go to infinity in the semiclassical limit. The semiclassical Rabi period, however, scales as $\tau_\text{R}\propto 1/(\lambda|\alpha|)=1/A$ and therefore remains fixed at a finite value. Additional timescales and modifications to the scaling behavior arise for different initial states and when the counter-rotating terms are included~\cite{Coleman2024, TwyeffortArmour2025}, but the overall picture remains valid: taking the semiclassical limit sends the collapse and revival times to infinity, while the semiclassical Rabi period remains fixed.

In the following  subsections, we present both numerical and analytical results for the convergence properties of each metric as $\lambda$ is decreased and $\lvert \alpha \rvert$ increased, keeping $\lambda|\alpha| = A$ constant. Numerical results are obtained by solving the time evolution under the quantum and semiclassical Hamiltonians with initial states of the form $\ket{+z}\otimes\ket{\alpha,n}$, where $\ket{+z}$ is the excited state of the spin. For the analytical results, we utilize the recently developed Floquet-basis rotating-wave approximation (FBRWA) solutions~\cite{TwyeffortArmour2025} that provide lowest-order quantum corrections to the semiclassical dynamics. We confine our analysis to the JCM, which allows for more tractable analytical approximations using the FBRWA. However, both the semiclassical limiting procedure and the FBRWA are valid for the full quantum Rabi model, so this approach will apply there as well and similar conclusions are expected.

\subsection{Trace distance}
We begin by quantifying the difference between quantum and semiclassical behavior for each of the subsystems using the trace distance~\cite{Breuer2009, Gilchrist2005, NielsenChuang} at a fixed instant in time, $t=t_1$.   
The trace distance between two density matrices $\rho$ and $\sigma$ is defined as 
\begin{equation}
    D(\rho , \sigma) \equiv \frac{1}{2} |\rho-\sigma |,
\end{equation}
where $|O| \equiv \sqrt{O^\dagger O}$. The quantum state of the coupled system is determined by numerically computing the evolution of the initial state $\ket{+z}\otimes\ket{\alpha,n}$ in the JCM. For the spin, the quantum state at $t_1$ is given by the reduced density matrix obtained by taking the partial trace over the field subsystem. The semiclassical state of the spin at $t_1$ is calculated using the corresponding semiclassical Hamiltonian
\begin{equation}
    \hat{H}_{sc}^{\text{RWA}}(t) = \frac{\Omega}{2}\hat{\sigma}_z + A(e^{i \omega_0 t} \hat{\sigma}_- + e^{-i \omega_0 t} \hat{\sigma}_+) ,
\end{equation}
where $A = \lambda \lvert \alpha \rvert$ with $\alpha$ assumed to be real, again taking the excited state $\ket{+z}$ as the initial state of the spin. The trace distance between the quantum and semiclassical states measures how closely the quantum state of the spin matches the semiclassical prediction at $t_1$.

The quantum density matrix for the field is likewise found by taking the partial trace of the coupled system's quantum state with respect to the spin. As discussed at the end of Sec.~\ref{sec:sc-limit}, in the semiclassical case the field is unaffected by the interaction with the spin. We therefore find the corresponding semiclassical state of the field by evolving the initial quantum field state under the bare Hamiltonian $\omega_0 \adag \a$. Again, the trace distance between the quantum and semiclassical states of the field serves as a measure of semiclassicality.

\begin{figure}[t]
  \centering
    \includegraphics[width=1\linewidth]{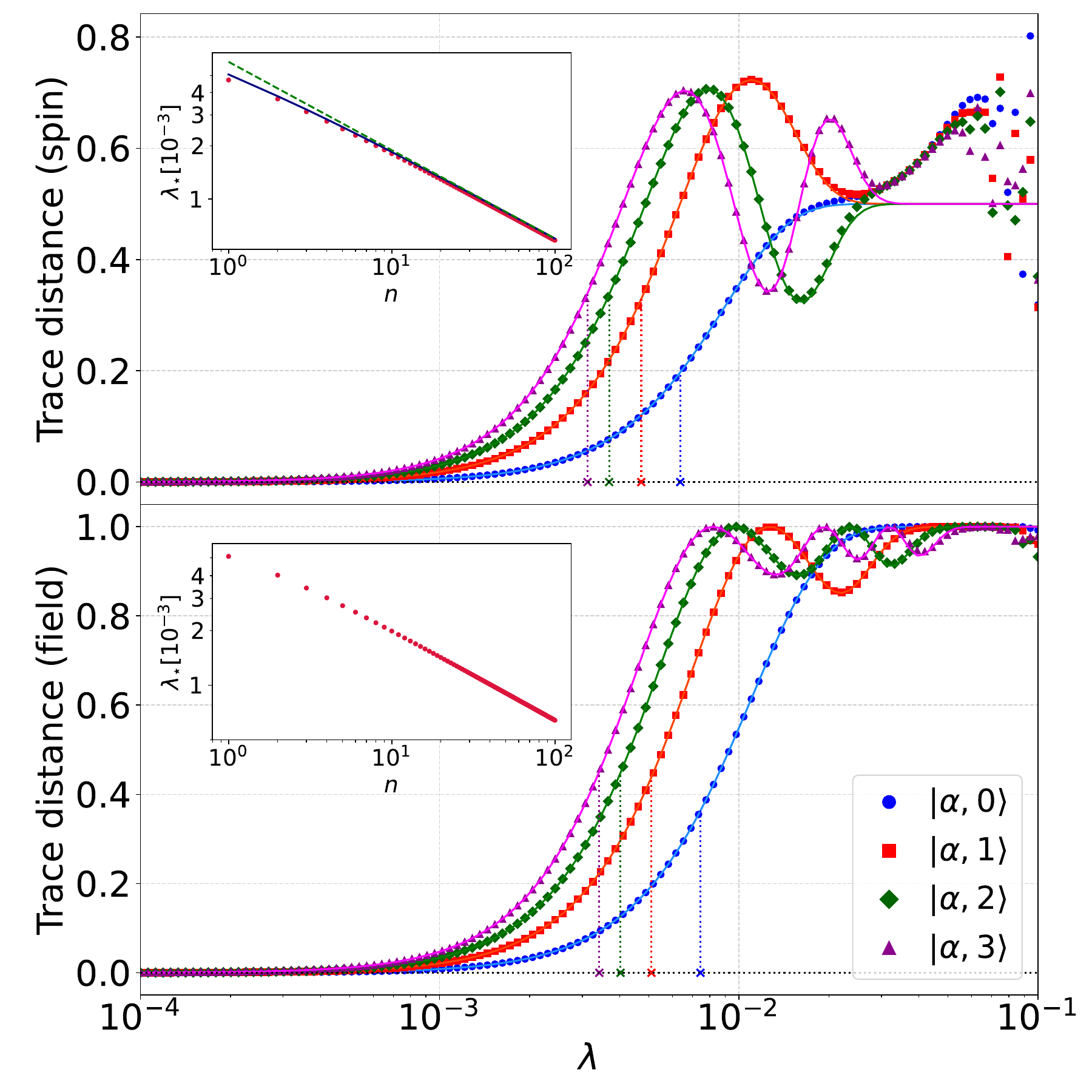}  
    \caption{Trace distance between quantum and semiclassical reduced density matrices for the spin (top panel) and field (bottom panel) at $t_1 = 10 \tau_\text{R}$ as a function of $\lambda$ (in units where $\omega_0 = 1$), with $A = \lambda \lvert \alpha \rvert = 0.2$. Initial states are of the form $\ket{+z}\otimes\ket{\alpha,n}$. Points indicate numerical solutions; solid curves show analytical results derived from the FBRWA. The colored dotted lines and crosses indicate the inflection points of the curves in the convergence region. Insets show the value of the coupling at the inflection point, $\lambda_\star$, as a function of $n$. Red points are numerical results; also shown for the spin are results from approximate analytical scaling relations based on a Taylor expansion (solid blue line) and its large-$n$ limit (dashed green line).}
    \label{fig:spin+field-td}
\end{figure}

In Fig.~\ref{fig:spin+field-td}, the trace distances for the spin (upper panel) and field (lower panel) are plotted at time $t=t_1=10 \tau_R$, where $\tau_R = 2A$ is the semiclassical Rabi period. The spin is initialized in its excited state $\ket{+z}$ and a range of displaced Fock states $\ket{\alpha,n}$ are considered for the initial state of the field. The coupling strength $\lambda$ is varied with $A=\lambda\lvert \alpha \rvert$ held constant. The results clearly show that the trace distances converge to zero in the limit of small $\lambda$ regardless of the initial field state. However, the convergence shifts to smaller coupling values as $n$ is increased, indicating that the quantum behavior remains more persistent as the limit is approached. 

Complementing the numerical results, analytical approximations are obtained from quantum-corrected Floquet dynamics within the FBRWA~\cite{TwyeffortArmour2025}. For an initially excited spin and a field initialized in a displaced Fock state $\ket{\alpha, n}$, the evolution under the JC Hamiltonian is approximately given by
\begin{equation}
\begin{split}
\ket{\Phi(t)}= & \tfrac{1}{2} e^{-i \omega_0 t (n + 1/2)} \\
& \times \Bigl[
e^{-i \lambda \lvert\alpha\rvert t / 2}
\left(\ket{+z}+e^{i \omega_0 t}\ket{-z}\right)\otimes\ket{\alpha_+(t),n}\\
&+\,e^{i \lambda \lvert\alpha\rvert t / 2}
\left(\ket{+z}-e^{i \omega_0 t}\ket{-z}\right)\otimes\ket{\alpha_-(t),n}\Bigr]
\label{eq:FBRWA_time_evolved_state}
\end{split}
\end{equation}
where $\ket{\alpha_\pm, n} = \hat{D} \bigl( |\alpha| e^{-i \omega_0 t}[1 \mp i \lambda t/ (2|\alpha|)] \bigr) \ket{n}$. Taking the partial trace over the field states yields
\begin{equation}
\rho_s^\text{FB}(t)=\frac{1}{2} \left[ \hat{I}_s + L_n (\lambda^2 t^2) e^{-\lambda^2 t^2 / 2} \, \sz \right].
\end{equation}
where subscripts $s$ and $f$ denote the spin and field subsystems, respectively.
With $t_1$ chosen to be an integer number of semiclassical Rabi periods $N \tau_\text{R}$, the trace distance for the spin becomes
\begin{equation}
D \bigl( \rho_s (t), \rho_s^\text{FB}(t) \bigr)
= \frac{1}{2}\left[ 1 - L_n(\lambda^2 t^2)\,e^{-\lambda^2 t^2/2} \right],
\label{eq:fbrwa-for-spin-td}
\end{equation}
where $t= N \tau_\text{R}$. This analytical expression is plotted alongside the numerical solutions in the top panel of Fig.~\ref{fig:spin+field-td}. The approximation captures the behavior of the trace distance remarkably well within the convergence region, deviating only at large values of $\lambda$. 

To study how varying $n$ impacts the emergence of semiclassical behavior,  Eq.~\eqref{eq:fbrwa-for-spin-td} is expanded in a Taylor series in $\lambda$: 
\begin{align}
    D \bigl( \rho_s (t), \rho_s^\text{FB}(t) \bigr) &= \frac{2n+1}{4} +  \lambda^2 t^2 \, - \, \frac{2n^2 +2n + 1}{16} \lambda^4 t^4  \nonumber \\
    &+  \frac{4n^3 +6n^2 + 8n + 3}{288} \lambda^6 t^6 + O(\lambda^8).
    \label{eq:taylor-expansion-for-sc-td}
\end{align}
Each of the trace distance curves plotted in Fig.~\ref{fig:spin+field-td} exhibits a similar form as $\lambda$ decreases. The inflection points indicated by vertical dotted lines serve as a convenient reference point on each curve. For couplings larger than this value, quantum corrections dominate the dynamics, whereas for smaller couplings the system exhibits a smooth onset of semiclassical behavior. The coupling at the inflection point, $\lambda_\star$, may be found from Eq.~\eqref{eq:taylor-expansion-for-sc-td}. In the large $n$ limit, it scales approximately as
\begin{equation}\label{eq:inflection-coupling-scaling-large-n}
    \lambda_\star \propto \frac{1}{t \sqrt{n}} \sqrt{\frac{9 -\sqrt{21}}{5}}.
\end{equation}
The inflection point couplings calculated using the Taylor expansion in Eq.~\eqref{eq:taylor-expansion-for-sc-td} and its large $n$ approximation, Eq.~\eqref{eq:inflection-coupling-scaling-large-n}, are shown in the inset within the top panel of Fig.~\ref{fig:spin+field-td} by the solid blue and dashed green lines, respectively. The red points in the inset indicate the inflection point couplings $\lambda_\star$ of each state $\ket{\alpha, n}$ extracted from the numerical solutions. The $1/\sqrt{n}$ scaling relationship holds well for all but the smallest few values of $n$. The gradient at the inflection point likewise scales as $\sqrt{n}$.

Corresponding analytical expressions for the field subsystem may also be calculated from the FBRWA state given in Eq.~\eqref{eq:FBRWA_time_evolved_state}. The partial trace over the spin states gives the reduced field density matrix
\begin{equation}
     \rho_f^{\text{FB}} = \frac{1}{2} \big[\ket{\alpha_+ (t) ,n}\bra{\alpha_+ (t) ,n} + \ket{\alpha_- (t) ,n}\bra{\alpha_- (t) ,n} \big].
\label{red. field density matrix for FBRWA}
\end{equation}
Note that $\rho_f^{\text{FB}}$ is an incoherent mixture of the two displaced number states $\ket{\alpha_\pm (t), n}$; the lack of coherences stems from the orthogonality of the Floquet basis states. 
The trace distance between the FBRWA state $\rho_f^{\text{FB}}(t)$ and the time-evolved field state $\rho_f(t)$ is given by
\begin{equation}
    D\bigl(\rho_f(t),\rho_f^{\text{FB}}(t)\bigr)
    = \sqrt{-\frac{a}{3}}\;\sum_{m=0}^{2}\bigl|\cos \bigl( \theta +2 \pi m/3\bigr)\bigr|,
    \label{eq:trace-distance}
\end{equation}
where $a$ and $b$ are real functions of overlaps between displaced Fock states, and
\begin{equation}
    \theta = \frac{1}{3}\arccos\!\left(\frac{-3b}{2a}\sqrt{-\frac{3}{a}}\right).
\end{equation}
Explicit expressions for $a$ and $b$, together with the derivation of Eq.~\eqref{eq:trace-distance}, are given in Appendix~\ref{app:td-calc}. The analytical prediction of Eq.~\eqref{eq:trace-distance} is plotted alongside the numerical results in the bottom panel of Fig.~\ref{fig:spin+field-td}. As for the spin, the approximate analytical results closely follow the numerical results in the vicinity of the semiclassical limit. The numerically calculated inflection-point couplings for the field trace distance, shown in the inset figure of the bottom panel of Fig.~\ref{fig:spin+field-td}, also approximately scale as $1/\sqrt{n}$. 

\subsection{Correlation}

The trace distance metrics capture the degree of semiclassicality of the spin and the field at a specified instant in time. Alternatively, we can consider the agreement between the quantum and semiclassical predictions over an extended time period. A measure based on the correlation between Fourier spectra was developed in recent work by two of the present authors to assess the temporal validity of the rotating-wave approximation~\cite{Coleman2024}. Here we adapt this metric for assessing the degree of semiclassicality in the spin population dynamics.

\begin{figure}[t]
    \centering
    
    \includegraphics[width=1\linewidth]{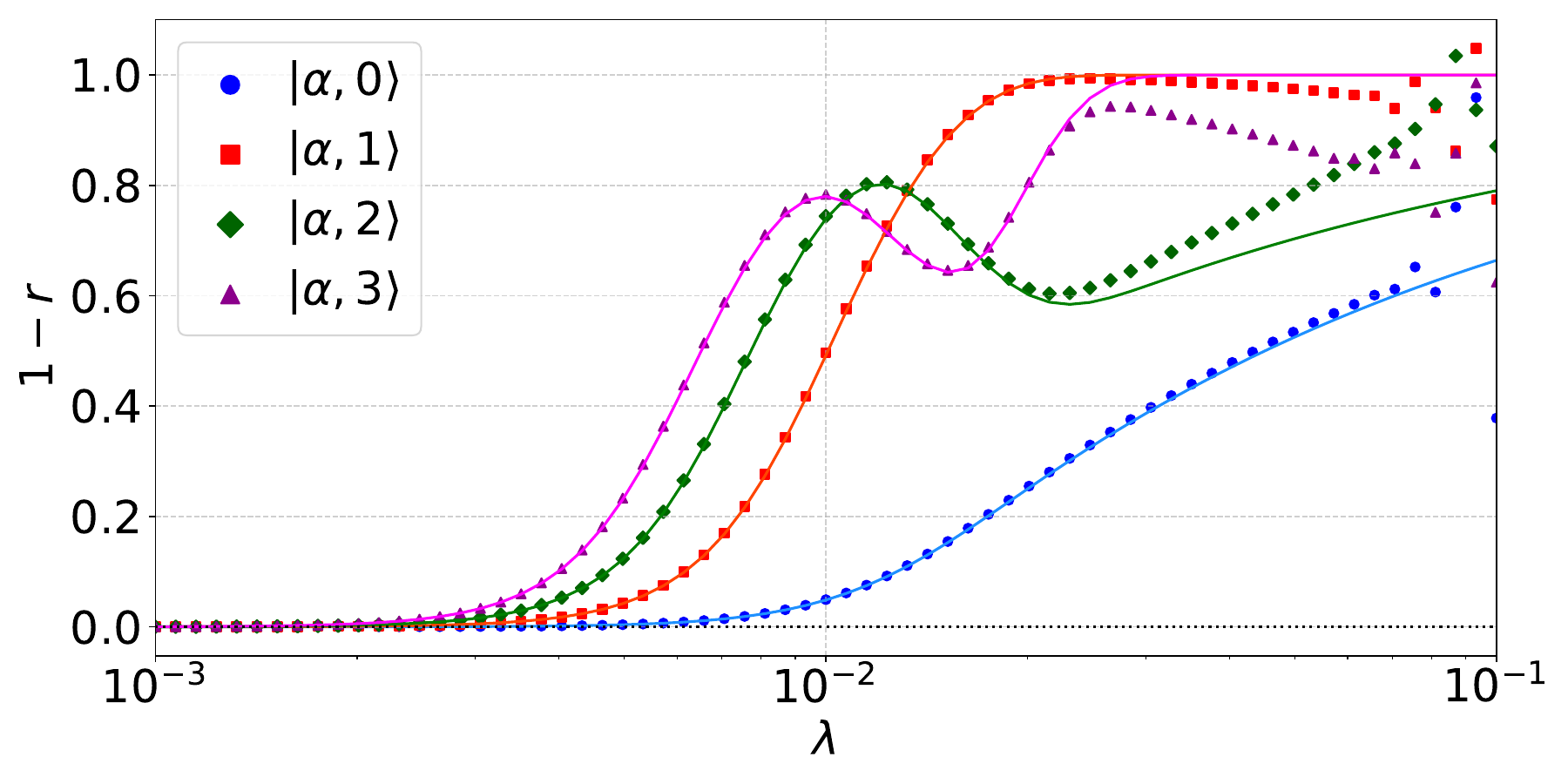}  

    \caption{Correlation between the Fourier spectra of the quantum and semiclassical atomic inversion dynamics over the period $0 \leq t \leq t_1$. The points and solid lines indicate the numerical and approximate analytical results, respectively. The parameters $t_1$ and $A=\lambda |\alpha|$ are the same as in Fig.~\ref{fig:spin+field-td}. Note the different scale on the $\lambda$ axis.}
    \label{fig:cos-sim}
\end{figure}

We begin by representing the spin dynamics in the frequency domain (for figures showing the spin excited-state population dynamics in semiclassical and quantum Rabi models and the corresponding Fourier transforms (FFTs) see Ref.~\cite{Coleman2024}). To construct a suitable measure, we compute the Pearson correlation coefficient $r$ between the FFTs of the quantum and semiclassical atomic inversions, where the atomic inversion is defined as $W(t) =\braket{\sz}$. For consistency, the corresponding metric is taken to be $1-r$ so that all measures converge to 0 as the semiclassical limit is approached. We refer to this quantity as the correlation. 
The atomic inversion is chosen in place of the excited-state population dynamics utilised in \cite{Coleman2024} for convenience when computing the corresponding analytical results. A full derivation of the analytical expression obtained within the FBRWA is presented in Appendix~\ref{app:cos-sim}. The numerical and analytical results for the correlation are shown in Fig.~\ref{fig:cos-sim}. Note the different scale on the horizontal axis compared to Fig.~\ref{fig:spin+field-td}, reflecting the more rapid convergence of the correlation measure as $\lambda$ is decreased. 

\subsection{Entanglement entropy}
Having considered the behavior of each subsystem individually, we now turn to the entanglement between them. In the semiclassical regime, we naturally expect no entanglement since only there is only one quantum subsystem. Although the semiclassical limiting procedure preserves the full bipartite Hilbert space structure, the form of Eq.~\eqref{eq:Hq-limit} prevents the generation of entanglement. To study the evolution of the entanglement as the limit is approached, we construct a measure based on the von Neumann entanglement entropy, widely employed as a standard measure of entanglement~\cite{Gerry2012-zd, PRXQuantum.5.030311}. We refer to this quantity simply as the entropy, defined as
\begin{equation}
S = - \tr [\rho \ln \rho],
\end{equation}
where $\rho$ is the reduced density operator. Noting that the entropies of the reduced states are equal, it is sufficient to work on the small spin subsystem~\cite{ArakiLieb1970}. 

To smooth out the oscillations of the entropy, the average is taken over a fixed number of semiclassical Rabi periods. 
Numerically, the average entropy is simple to calculate; analytically, the entropy must be integrated over some time window. The analytical calculation using the FBRWA is detailed in Appendix~\ref{app:vn-entropy}. Figure~\ref{fig:average-vn-entropy} shows the numerical and analytical results for the average entropy. The dashed line at $\overline{S}(\lambda)= \ln 2$ indicates the maximum possible average entropy. Again, the results demonstrate convergence to semiclassical behavior following a similar pattern to the previous measures. The agreement between the numerical solutions and the analytical approximation is excellent within the convergence region~\footnote{The keen eye may notice that the analytical results for $\Bar{S}$ sit slightly above 0 in the small coupling regime. Truncation of the infinite series contained in Eq.~\eqref{eq:avgS_gen_final} causes slight inaccuracies in the numerical evaluation.}. However, the analytical FBRWA prediction breaks down at larger couplings, with all solid curves converging to $\ln 2$. This has an intuitive interpretation. Within the approximation, the two field components in Eq.~\eqref{eq:FBRWA_time_evolved_state} become displaced in phase space in opposite directions. As time goes on, the displaced field states become distinguishable as their overlap decreases, causing the subsystems to entangle. The FBRWA does not capture the revival behavior, instead predicting a linear increase in the field-state displacements with time. Hence the system tends toward maximum entanglement. 
\begin{figure}[t]
    \centering
    
    \includegraphics[width=1\linewidth]{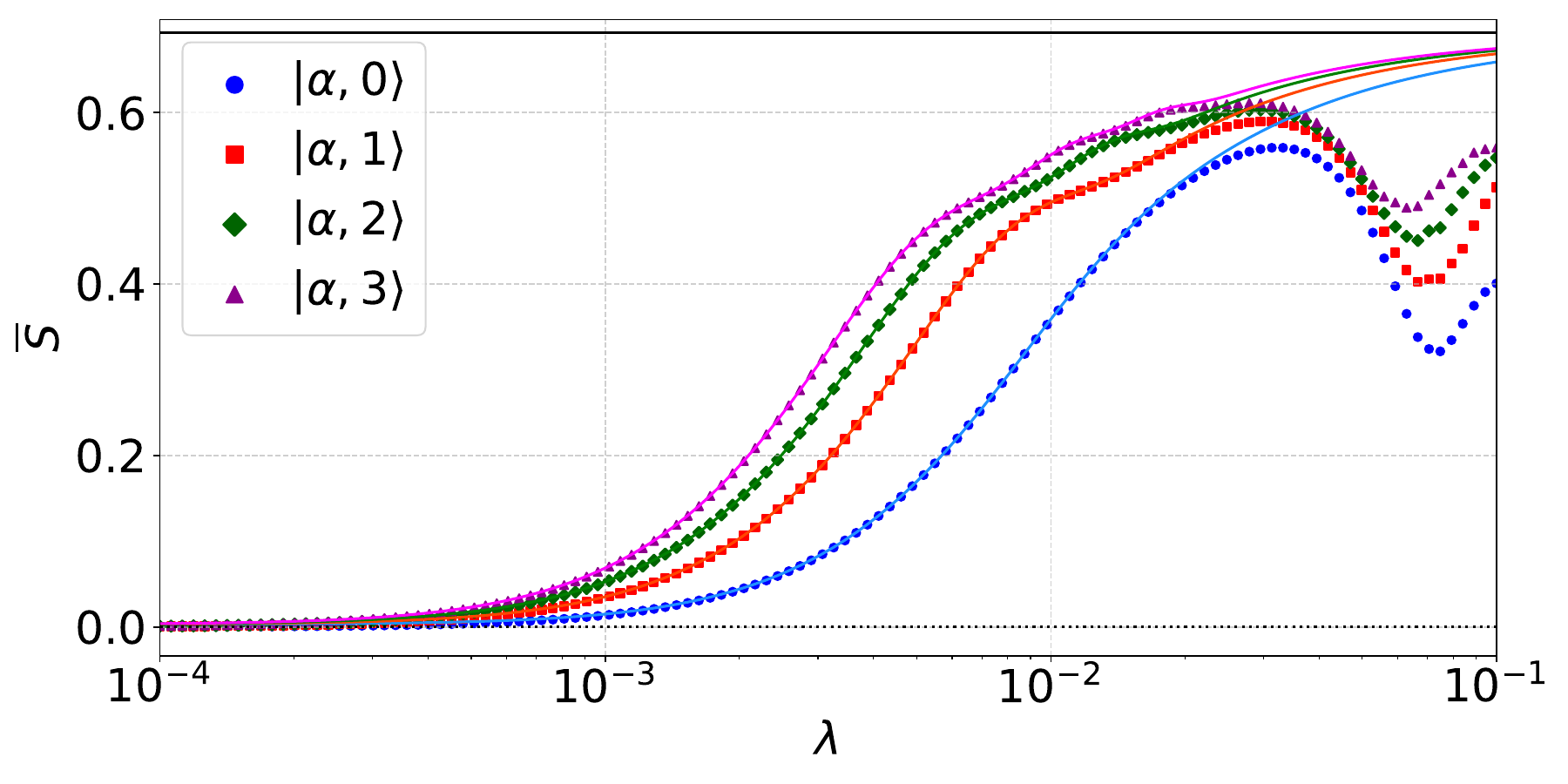}  

    \caption{Average von Neumann entropy over an integer number of Rabi periods. A horizontal line at $\overline{S}(\lambda)=\ln2$ indicates the maximum possible value. The points and solid lines indicate the numerical and approximate analytical results, respectively; $t_1$ and $A=\lambda |\alpha|$ are the same as in the previous figures.}
    \label{fig:average-vn-entropy}
\end{figure}

\subsection{Discussion}

The range of measures considered here generate a consistent set of predictions for the emergence of semiclassical behavior in the limit $\lambda \to 0$ and $\lvert \alpha \rvert \to \infty$. Notably, semiclassical dynamics of the spin does not depend on the field being in a coherent state. Both numerical and analytical results support the assertion of \cite{Irish2022} that any displaced Fock state $\ket{\alpha, n}$ with the same value of $\alpha$ gives rise to identical semiclassical spin dynamics in the limit. However, the convergence properties depend on $n$ --- as $n$ increases, it is necessary to go deeper into the limit to obtain the same degree of semiclassicality. Analytical scaling relationships obtained for the trace distance measures indicate that the value of $\lambda$ at which monotonic convergence to the limit sets in goes as $n^{-1/2}$.

These results demonstrate excellent agreement between the analytical approximations obtained using the FBRWA and the numerical computations in the convergence region. For all four measures, the approximations break down at broadly similar values of $\lambda$. This breakdown is expected since the validity of the approximation improves as phase-space displacement of the initial state is increased \cite{TwyeffortArmour2025}; as $\lambda |\alpha|$ is a fixed quantity, this translates to  an improvement in the approximation as $\lambda$ is decreased.

While the results plotted in Figs. 1-3 are calculated for an evolution time of $t_1 = 10 \tau_R$, the conclusions are independent of the choice of $t_1$. Extending the windowed dynamics increases sensitivity to collapse phenomena; at fixed drive amplitude $A$, the coupling must be reduced so that the collapse remains negligible over the longer timescale. Similarly, the analytical expression given in Eq.~\eqref{eq:inflection-coupling-scaling-large-n} shows that the inflection point coupling is inversely proportional to $t_1$. Therefore, increasing the evolution time simply shifts the convergence to smaller values of the coupling. This has an intuitive explanation in the QCFD framework, in which terms involving quantum field operators appear as perturbations to the semiclassical spin dynamics, characterized by the quantum coupling $\lambda$. 


\section{Conclusions}\label{sec:discussion}

We have investigated how semiclassical behavior emerges from the quantum Rabi model by means of the rigorous limiting procedure introduced in Ref.~\cite{Irish2022}. While that procedure establishes correspondence at the Hamiltonian level, here we have examined convergence for specific quantum states, using displaced Fock states as a natural basis. The approach to semiclassical dynamics has been quantified through measures that capture three distinct aspects of semiclassical behavior: the trace distance of reduced states, the correlation of spin population inversion dynamics, and the average von Neumann entanglement entropy. For each of these measures, numerical computations clearly demonstrate that all states in the set of displaced Fock states give rise to the same semiclassical dynamics in the joint limit $\lambda \to 0$ and $|\alpha|\to\infty$ with the product $\lambda|\alpha |$ held fixed. Our analysis further reveals that the rate of convergence depends on the initial field state, with quantum behavior persisting further into the limit  for displaced Fock states with larger $n$. The FBRWA, which accounts for corrections to the semiclassical dynamics arising from field quantization, has been used to derive  analytical approximations that capture the qualitative features and the $1/\sqrt{n}$ scaling relation between the coupling at which the inflection point of convergence and the initial field state $\ket{\alpha, n}$.

The results we have presented provide strong support for the mathematical formulation of the semiclassical limit proposed in \cite{Irish2022}. Moreover, they further establish the power and utility of the QCFD approach to calculating quantum field corrections to the semiclassical solutions and the accuracy of the closed-form approximate expressions obtained from the FBRWA.~\cite{TwyeffortArmour2025} These findings offer fresh insight into the fundamental question of how the semiclassical dynamics of a quantum system arises from its interaction with a quantum field.

\section*{Data Availability}
The Python code that supports the findings of this manuscript is openly
available~\cite{github}.

\section*{Acknowledgments}

We thank Mason Spittle and Jonas Glatthard for helpful conversations. ADA acknowledges support from a Leverhulme Trust Research Project Grant (RPG-2023-177).
\appendix

\section{Derivation of the field trace distance}\label{app:td-calc}
In this Appendix, we derive the analytical form of the trace distance between the time-evolved reduced field states of the semiclassical Hamiltonian in Eq.~\eqref{eq:Hq-limit} and the effective FBRWA state. 

The trace distance between the time-evolved reduced field state
$\rho_f$ and the effective FBRWA state $\rho_f^{\mathrm{FB}}$ is given by
\begin{equation}
  D(\rho_f^{\mathrm{FB}},\rho_f)
  = \frac12 \lVert \Delta \rVert_1
\end{equation}
where $ \Delta \equiv \rho_f^{\mathrm{FB}} - \rho_f$.
These states are given by the outer products
\begin{equation}
  \rho_f^{\mathrm{FB}}
  = \tfrac12\bigl(\ket{\alpha_+}\bra{\alpha_+}
                   + \ket{\alpha_-}\bra{\alpha_-}\bigr),
\end{equation}
and
\begin{equation}
  \rho_f = \ket{\alpha_0}\bra{\alpha_0},
\end{equation}
with
\begin{align}
  \ket{\alpha_0}
  &\equiv \hat{D}(\alpha e^{-i\omega_0 t})\ket{n},\\
  \ket{\alpha_\pm}
  &\equiv \hat{D}\!\left(\alpha e^{-i\omega_0 t}
           \Bigl[1 \mp i\lambda t/(2|\alpha|)\Bigr]\right)\ket{n}.
\end{align}
Therefore, $\Delta$ can be written as a linear combination of three rank-1
projectors. By construction, $\Delta$ has support contained in the subspace
\begin{equation}
  \mathcal{S}
  = \mathrm{span}\{\ket{\alpha_0},\ket{\alpha_+},\ket{\alpha_-}\}.
\end{equation}
In particular, $\Delta$ acts trivially on the orthogonal complement
$\mathcal{S}^\perp$ and can therefore be represented on, at most, a three-dimensional subspace $\mathcal{S}$.
To compute the trace distance, it suffices to determine the
eigenvalues of this $3\times3$ representation. We first record the
overlaps between the kets in $\mathcal{S}$.
One finds
\begin{align}
\braket{\alpha_0|\alpha_\pm}
&= e^{-\lambda^2 t^2/8}\,L_n\bigl(\lambda^2 t^2/4\bigr)\,e^{\mp i\phi/2}
   \equiv \kappa\,e^{\mp i\phi/2}
   \equiv c_{0,\pm}, \\
\braket{\alpha_+|\alpha_-}
&= e^{-\lambda^2 t^2/2}\,L_n\bigl(\lambda^2 t^2\bigr)\,e^{i\phi}
   \equiv \zeta\,e^{i\phi},
\end{align}
where $\phi = \lambda\,|\alpha|\,t$ and, in particular,
$\kappa \in \mathbb{R}$.

Starting from the nonorthogonal set
\(\mathcal{S} = \{\ket{\alpha_0},\ket{\alpha_+},\ket{\alpha_-}\}\),
we construct an orthonormal basis via the Gram–Schmidt procedure. We
first set
\begin{equation}
  \ket{1} = \ket{\alpha_0},
\end{equation}
which is already normalized. The second basis vector is
\begin{equation}
  \ket{2}
  = \frac{\ket{\alpha_+} - \braket{1|\alpha_+}\ket{1}}
         {\sqrt{1-|\braket{1|\alpha_+}|^2}}
  = \frac{\ket{\alpha_+} - c_{0,+}\ket{1}}{\sqrt{1-\kappa^2}},
\end{equation}
using $|c_{0,+}|^2 = \kappa^2$. The third basis vector is
\begin{equation}
  \ket{3}
  = \frac{\ket{\alpha_-} - \braket{1|\alpha_-}\ket{1}
                       - \braket{2|\alpha_-}\ket{2}}
         {\sqrt{1-\kappa^2 - |\braket{2|\alpha_-}|^2}}.
\end{equation}
In this orthonormal basis $\{\ket{1},\ket{2},\ket{3}\}$ we introduce
the shorthand coefficients
\begin{equation}
  c_{0,\pm} = \braket{1|\alpha_\pm},\quad
  c_{1,\pm} = \braket{2|\alpha_\pm},\quad
  c_{2,-}   = \braket{3|\alpha_-},
\end{equation}
so that
\begin{align}
  \ket{\alpha_+} &= c_{0,+}\ket{1} + c_{1,+}\ket{2}, \\
  \ket{\alpha_-} &= c_{0,-}\ket{1} + c_{1,-}\ket{2} + c_{2,-}\ket{3}.
\end{align}

In the orthonormal basis $\{ \ket{1}, \ket{2}, \ket{3} \}$, the
operator $\Delta$ is represented by
\begin{widetext}
\begin{equation}
\Delta \;=\;
\begin{bmatrix}
\kappa^2 - 1
&
\tfrac12\bigl(c_{0,+}\,c_{1,+}^* \;+\; c_{0,-}\,c_{1,-}^*\bigr)
&
\tfrac12\,c_{0,-}\,c_{2,-}^*
\\[1ex]
\tfrac12\bigl(c_{0,+}^*\,c_{1,+} \;+\; c_{0,-}^*\,c_{1,-}\bigr)
&
\tfrac12\bigl((1-\kappa^2) \;+\; |c_{1,-}|^2\bigr)
&
\tfrac12\,c_{1,-}\,c_{2,-}^*
\\[1ex]
\tfrac12\,c_{0,-}^*\,c_{2,-}
&
\tfrac12\,c_{1,-}^*\,c_{2,-}
&
\tfrac12\bigl(1 - \kappa^2 - |c_{1,-}|^2\bigr)
\end{bmatrix}.
\end{equation}
\end{widetext}
We denote the matrix elements of $\Delta$ in this basis by
$\Delta_{ij}$, $i,j=0,1,2$.
Since $\Delta$ is the difference of two density operators, it is
Hermitian and its eigenvalues are real. The characteristic polynomial
$\det(\Delta - \chi \hat{I})$ can be written as
\begin{equation}
  \chi^3 + a\,\chi + b = 0,
\end{equation}
because $\tr\Delta = 0$ and hence the quadratic term vanishes. The
coefficient $a$ is given by
\begin{multline}
  a
  = \Delta_{00}\Delta_{11}+\Delta_{00}\Delta_{22}+\Delta_{11}\Delta_{22} \\
    - \bigl(|\Delta_{01}|^2+|\Delta_{02}|^2+|\Delta_{12}|^2\bigr),
\end{multline}
while
\begin{equation}
  b = -\det\Delta.
\end{equation}
The three eigenvalues $\chi_k$ ($k=0,1,2$) of this depressed cubic
can be expressed in trigonometric form as
\begin{equation}
  \chi_k = R\,\cos\bigl(\theta + 2\pi k/3\bigr),
  \qquad k=0,1,2,
\end{equation}
where
\begin{equation}
  R = 2\sqrt{-\frac{a}{3}},
  \qquad
  \theta = \frac{1}{3} \arccos\! \left(\frac{-3b}{2a}\sqrt{-\frac{3}{a}}\right).
\end{equation}

Finally, the trace distance between $\rho_f^{\mathrm{FB}}$ and
$\rho_f$ is obtained by summing the absolute values of these
eigenvalues:
\begin{equation}
  D(\rho_f^{\mathrm{FB}},\rho_f)
  = \frac12\sum_{k=0}^2 |\chi_k|.
\end{equation} 

\onecolumngrid 

\section{Pearson correlation coefficient}\label{app:cos-sim}

In this Appendix we calculate the Pearson correlation coefficient between the Fourier transforms of the JC and semiclassical RWA atomic inversions, $W(t)$, to measure the emergence of semiclassical behavior over the entire dynamics~\cite{Coleman2024}. Applying the FBRWA to the JC model one finds
\begin{equation}
W^{\text{FB}}(t) = e^{-\tfrac{1}{2}\lambda^2 t^2}\, L_n(\lambda^2 t^2)\,\cos(2\lambda|\alpha|t).   
\end{equation}
The atomic inversion in the semiclassical RWA model is
\begin{equation}
W_{\text{scRWA}}(t) = \cos(2\lambda|\alpha|t),
\end{equation}
In what follows, we restrict our attention to a finite window of dynamics, $0 \leq t \leq T$. We work with the atomic inversion rather than the excited-state population, since the inversion has a negligible mean over the dynamics.

In the case of continuous functions, the correlation between two functions $f$ and $g$, whose means are 0, can be defined as 
\begin{equation}
r(f, g) =
\frac{ \int_{-\infty}^{\infty} \tilde{f}(\omega)\,\tilde{g}(\omega)^* \, \mathrm{d}\omega}
{ \sqrt{\int_{-\infty}^{\infty} |\tilde{f}(\omega)|^2 \,\mathrm{d}\omega}\;
\sqrt{\int_{-\infty}^{\infty} |\tilde{g}(\omega)|^2 \,\mathrm{d}\omega}} = \frac{ \int_{-\infty}^{\infty} f(t)\, g(t)^* \, \mathrm{d}t}
{ \sqrt{\int_{-\infty}^{\infty} |f(t)|^2 \,\mathrm{d}t}\;
\sqrt{\int_{-\infty}^{\infty} |g(t)|^2 \,\mathrm{d}t}}.
\label{eq:def-cossim}
\end{equation}
where $\tilde{g}(\omega) $ and $\tilde{f}(\omega)$ are the Fourier transforms of $g(t)$ and $f(t)$, respectively. In Eq.~\eqref{eq:def-cossim}, we have used the Plancherel theorem to express the correlation in terms of time-domain inner products. Thus the correlation may be viewed either as a normalized inner product in frequency space or, equivalently by Plancherel's theorem, as a normalized time-domain inner product. In practice, we evaluate this correlation over a finite time window by replacing the integrals over $(-\infty,\infty)$ with $\int_0^T$.
Over the time window $[0, T]$, the numerator is given by
\begin{align}
\int_0^T   W^{\text{FB}}(t)\,W_{\text{scRWA}}(t) \, \mathrm{d}t
&=\int_0^T \exp[-\lambda^2 t^2 /2] L_n (\lambda^2 t^2) \cos^2(2 \lambda |\alpha |t) \,\mathrm{d}t \nonumber \\
&= \frac{1}{2}\sum_{r=0}^n a_{n,r}\,\lambda^{2r}
\Bigg[
{\int_0^T t^{2r}e^{- \mu t^2}\,\mathrm{d}t}
+{\int_0^T t^{2r}e^{- \mu t^2}\cos(2 \nu t)\,\mathrm{d}t}
\Bigg], \label{eq:cos-sim-numerator-integral}
\end{align}
where $\nu = 2\lambda |\alpha|$ and $\mu = \lambda^2/2$; we have written the Laguerre polynomial in summation form $L_n(x)=\sum_{r=0}^n a_{n,r}\,x^r$, where $a_{n, r} = \binom{n}{r}\,(-1)^r/{r!}$ and used the trigonometric identity $2\cos^2 x = \cos2x + 1$. For the non-oscillatory term, we use the change of variables $u=x t^2$
\begin{align}
\int_0^T t^{2r} e^{-x t^2}\,\mathrm{d}t
&= \int_{0}^{x T^{2}} \left(\frac{u}{x}\right)^{r} e^{-u}\, \nonumber
\frac{1}{2\sqrt{x}}\,u^{-1/2}\,\mathrm{d}u \\
&= \frac{1}{2}\,x^{-(r+\frac12)} \int_{0}^{x T^{2}} u^{\,r-\frac12}\,e^{-u}\,\mathrm{d}u \nonumber \\
&= \frac{1}{2}\,x^{-(r+\frac12)} \int_{0}^{x T^{2}} u^{\,(r+\frac12)-1}\,e^{-u}\,\mathrm{d}u.
\end{align}
Identifying the lower incomplete gamma function;
\begin{equation}
\int_0^T t^{2r} e^{-x t^2}\,\mathrm{d}t
= \frac{1}{2}\,x^{-(r+\frac12)}\,\gamma\Big(r+\tfrac12,\;x T^2\Big).
\label{eq:gamma-moment}
\end{equation}
where $\gamma(s,z)$ is the lower incomplete gamma function. The oscillatory term in Eq.~\eqref{eq:cos-sim-numerator-integral} may be rewritten as 
\begin{equation}
\int_0^T t^{2r} e^{- \mu t^2}\cos(2 \nu t)\,\mathrm{d}t = \Re \left[ (-1)^r \, \partial_\mu^r \int_0^T e^{- \mu t^2} e^{-i 2\nu t} \,\mathrm{d}t \right]
\end{equation}
where we have used the fact $\partial_\mu I = \int_0^T (-t^2)e^{-\mu t^2}e^{-i2\nu t}\,\mathrm{d}t$, where $I$ is defined as
\begin{equation}
I(\mu,\nu;T)\;=\;\int_0^T e^{-\mu t^2}\,e^{-i\nu t}\,\mathrm{d}t = e^{-\nu^2/(4\mu)}\int_0^T
e^{-\mu\left(t + \frac{i\nu}{2\mu}\right)^2} \,\mathrm{d}t,
\qquad \mu>0,\; \nu\in\mathbb{R},
\end{equation}
Making the change of variable $u = \sqrt{\mu}\,t \;+\; i\,\frac{\nu}{2\sqrt{\mu}}$ the integral may be rewritten as 
\begin{equation}\label{eq:error-func-integral}
   I(\mu,\nu;T)
= e^{-\nu^2/(4\mu)}\;\frac{1}{\sqrt{\mu}}
        \int_{i\,\frac{\nu}{2\sqrt{\mu}}}^{\sqrt{\mu}\,T + i\,\frac{\nu}{2\sqrt{\mu}}} e^{-u^2}\,\mathrm{d}u.
\end{equation}
Then using the definition of the error function $\erf(z)$, Eq.~\eqref{eq:error-func-integral} can be rewritten as
\begin{equation}\label{eq:I-func-cos-sim}
I(\mu,\nu;T)
= \frac{\sqrt{\pi}}{2\sqrt{\mu}}\,
e^{-\nu^2/(4\mu)}\!
\left[
\erf \Big(\sqrt{\mu}\,T + i\,\frac{\nu}{2\sqrt{\mu}}\Big)
- \erf\Big(i\,\frac{\nu}{2\sqrt{\mu}}\Big)
\right].
\end{equation}
Therefore, Eq.~\eqref{eq:cos-sim-numerator-integral} takes the form
\begin{equation}\label{eq:cos-sim-numerator}
\int_0^T W^{\text{FB}}(t)\,W_{\text{scRWA}}(t)\,\,\mathrm{d}t
=\frac{1}{2}\sum_{r=0}^n a_{n,r}\,\lambda^{2r}
\bigg[
\frac{1}{2}\,\mu^{-(r+\frac12)}\gamma \Big(r+\tfrac12,\mu T^2\Big)
+ \Re \Big((-1)^r\partial_\mu^{\,r} I(\mu, 2\nu;T)\Big)
\bigg].
\end{equation}

The denominator terms can be calculated in a similar way
\begin{align}
\int_0^T W^{\text{FB}}(t)^2\,\mathrm{d}t
&=\int_0^T e^{-2 \mu t^2}\,\big[L_n(\lambda^2 t^2)\big]^2\,\cos^2(\nu t)\,\mathrm{d}t \nonumber \\
&= \frac{1}{2}\sum_{j=0}^{2n} b_{n,j}\,\lambda^{2j}
\left[
\int_0^T t^{2j} e^{-2 \mu t^2}\,\mathrm{d}t
+
\int_0^T t^{2j} e^{-2 \mu t^2}\cos(2 \nu t)\,\mathrm{d}t
\right].
\end{align}
where we have written $L_n (\lambda ^2 t^2)^2 = \sum_{j=0}^{2n} b_{n,j}\,(\lambda^2 t^2)^j$, where $b_{n,j}=\sum_{r=\max(0,j-n )}^{\min(n, j)} a_{n,r}\,a_{n,j-r}$.
Therefore
\begin{equation}\label{eq:norm-gn}
\int_0^T W^{\text{FB}}(t)^2\,\mathrm{d}t
=\frac{1}{2}\sum_{j=0}^{2n} b_{n,j}\,\lambda^{2j}
\bigg[
\frac{1}{2}\,(2\mu)^{-(j+\frac12)}\gamma \Big(j+\tfrac12,2 \mu T^2\Big)
+ \Re \Big((-1)^j\partial_{(2 \mu)}^{\,j} I(2\mu , 2\nu ;T)\Big)
\bigg]
\end{equation}

One may easily confirm that $\int_0^T W_{\text{scRWA}}(t)^2 = T/2$ when $T$ is an integer multiple of the semiclassical Rabi period. Combining \eqref{eq:cos-sim-numerator}, and \eqref{eq:norm-gn}:
\begin{equation}
r^n
=\frac{\displaystyle
\frac{1}{2} \sum_{r=0}^n a_{n,r}\,\lambda^{2r}
\left[
\frac{1}{2}\,\mu^{-(r+\frac12)}\gamma\big(r+\tfrac12,\mu T^2\big)
+ \Re \big((-1)^r\partial_\mu^{\,r} I(\mu,2\nu;T)\big)
\right]}
{\displaystyle
\sqrt{\frac{T}{4} \sum_{j=0}^{2n} b_{n,j}\,\lambda^{2j}
\left[
\frac{1}{2}\,(2\mu)^{-(j+\frac12)}\gamma \big(j+\tfrac12,2 \mu T^2\big)
+ \Re \big((-1)^j\partial_{(2\mu)}^{\,j} I(2\mu,2\nu;T)\big)
\right]}}
\label{eq:final-general-n}
\end{equation}
with $\mu=\lambda^2/2$, $\nu=2\lambda|\alpha|$, $T=N\pi/(\lambda|\alpha|)$, the function $I$ is given explicitly in \eqref{eq:I-func-cos-sim}.

\twocolumngrid 
\onecolumngrid 

\section{Averaged von Neumann entropy over $N$ Rabi periods}\label{app:vn-entropy}

Here we show the derivation of the averaged von Neumann entropy generated between the subsystems over the evolution, for arbitrary $n\in\mathbb N$. Defining
\begin{equation}\label{eq:X_def}
X(t) \equiv L_n(\lambda^2 t^2)\,e^{-\lambda^2 t^2/2},
\end{equation}
the diagonal entries may be written as
\begin{equation}\label{eq:rho_diag_gen}
\rho_{\pm\pm}(t)=\frac{1}{2}\left(1\pm X(t)\right),
\end{equation}
and the off-diagonal elements are 0. The instantaneous von Neumann entropy is then
\begin{align}\label{eq:S_bin_gen}
S(t)&=- \rho_{++}(t)\ln\rho_{++}(t) - \rho_{--}(t)\ln\rho_{--}  \nonumber \\
 &=-\frac{1+X(t)}{2}\ln\!\left(\frac{1+X(t)}{2}\right)
-\frac{1-X(t)}{2}\ln\!\left(\frac{1-X(t)}{2}\right).
\end{align}
To average the entropy, we integrate over a fixed number of semiclassical Rabi periods
\begin{equation}\label{eq:avgS_def_N}
\overline S^{(N)}=\frac{1}{N\tau_{\text{R}}}\int_{0}^{N\tau_{\text{R}}} S(t)\,\mathrm dt.
\end{equation}
Let $X\in(-1,1)$. Differentiating Eq.~\eqref{eq:S_bin_gen}
\begin{equation}\label{eq:dS_dX}
\frac{d}{dX}\Biggl[
-\frac{1+X}{2}\ln\!\left(\frac{1+X}{2}\right)
-\frac{1-X}{2}\ln\!\left(\frac{1-X}{2}\right)
\Biggr]
=\frac{1}{2}\ln\!\left(\frac{1-X}{1+X}\right).
\end{equation}
Using $\tanh^{-1}X=\sum_{k=0}^{\infty}\frac{X^{2k+1}}{2k+1}$ for $|X|<1$ and
$\ln\bigl(\tfrac{1+X}{1-X}\bigr)=2\,\tanh^{-1}X$, Eq.~\eqref{eq:dS_dX} becomes
\begin{equation}\label{eq:dS_series}
\frac{d}{dX}S=\frac{1}{2}\ln\!\left(\frac{1-X}{1+X}\right)
=-\sum_{k=0}^{\infty}\frac{X^{2k+1}}{2k+1},\qquad |X|<1,
\end{equation}
where we have dropped the explicit time dependence. Integrating from $0$ to $X$,
\begin{equation}
S(X)-S(0)
=-\sum_{k=0}^{\infty}\int_{0}^{X}\frac{u^{2k+1}}{2k+1}\,\mathrm{d}u
=-\sum_{k=0}^{\infty}\frac{X^{2k+2}}{(2k+1)(2k+2)}.
\end{equation}
Since $S(0)= -\tfrac12\ln\tfrac12 - \tfrac12\ln\tfrac12=\ln2$, and changing the summation $m=k+1$, we get
\begin{equation}\label{eq:S_series_time}
S(t)=\ln 2-\sum_{m=1}^{\infty}\frac{\bigl[L_n(\lambda^2 t^2)\bigr]^{2m}\,e^{-m\lambda^2 t^2}}{2m(2m-1)}.
\end{equation}

Averaging as in Eq.~\eqref{eq:avgS_def_N} one finds
\begin{equation}\label{eq:avgS_gen_start}
\overline S^{(N)}=\ln 2-\frac{1}{N\tau_{\text{R}}}
\sum_{m=1}^{\infty}\frac{1}{2m(2m-1)}
\int_{0}^{N\tau_{\text{R}}}\bigl[L_n(\lambda^2 t^2)\bigr]^{2m}e^{-m\lambda^2 t^2}\,\mathrm dt.
\end{equation}
Introducing the dimensionless upper limit $\chi_N$ and using the change variables $x=\lambda t$, then $s=x^2$,
\begin{align}
\int_{0}^{N\tau_{\text{R}}}\bigl[L_n(\lambda^2 t^2)\bigr]^{2m}e^{-m\lambda^2 t^2}\,\mathrm dt
&=\frac{1}{\lambda}\int_{0}^{\chi_N}\bigl[L_n(x^2)\bigr]^{2m}e^{-m x^2}\,\mathrm dx \nonumber\\
&=\frac{1}{2\lambda}\int_{0}^{\chi_N^2}\bigl[L_n(s)\bigr]^{2m}\,s^{-1/2}e^{-m s}\,\mathrm ds.
\label{eq:time_int_dimless}
\end{align}
Next we write $L_n$ as a summation,
\begin{equation}\label{eq:Ln_power}
L_n(s)=\sum_{r=0}^{n}a_r s^r,\qquad
a_r=\frac{(-1)^r}{r!}\binom{n}{r},
\end{equation}
and raise to the $2m$-th power,
\begin{equation}\label{eq:Ln_power_2m}
\bigl[L_n(s)\bigr]^{2m}=\sum_{\ell=0}^{2mn}c_{n,m,\ell}\,s^{\ell},\qquad
c_{n,m,\ell}=\sum_{\substack{r_1+\cdots+r_{2m}=\ell\\ 0\le r_i\le n}}\prod_{i=1}^{2m}a_{r_i}.
\end{equation}
Starting from \eqref{eq:time_int_dimless} and inserting
$\bigl[L_n(s)\bigr]^{2m}=\sum_{\ell=0}^{2mn} c_{n,m,\ell}\,s^{\ell}$, we obtain
\begin{equation}
\frac{1}{2\lambda}\sum_{\ell=0}^{2mn} c_{n,m,\ell}
\int_{0}^{\chi_N^2} s^{\ell-\frac12} e^{-m s}\,\mathrm ds.
\end{equation}
With the substitution $u=ms$ (so $s=u/m$ and $ds=du/m$), the integral becomes
\begin{align}
\int_{0}^{\chi_N^2} s^{\ell-\frac12} e^{-m s}\,ds
&=\int_{0}^{m\chi_N^2} \left(\frac{u}{m}\right)^{\ell-\frac12} e^{-u}\,\frac{\mathrm du}{m} \\
&= m^{-\ell-\frac12}\int_{0}^{m\chi_N^2} u^{\ell-\frac12} e^{-u}\,\mathrm du
= m^{-\ell-\frac12}\,\gamma\, \left(\ell+\tfrac12,\;m\chi_N^2\right),
\end{align}
which yields \eqref{eq:gamma_piece} and hence \eqref{eq:time_int_final}.
\begin{equation}\label{eq:gamma_piece}
\int_{0}^{\chi_N^2}\! s^{\ell-\frac12}e^{-m s}\,\mathrm ds
=m^{-\ell-\frac12}\,\gamma\!\left(\ell+\tfrac12,\,m \chi_N^2\right),
\end{equation}
with $\gamma(s,z)$ the lower incomplete gamma function. Hence
\begin{equation}\label{eq:time_int_final}
\int_{0}^{N\tau_{\text{R}}}\bigl[L_n(\lambda^2 t^2)\bigr]^{2m}e^{-m\lambda^2 t^2}\,\mathrm dt
=\frac{1}{2\lambda}\sum_{\ell=0}^{2mn}c_{n,m,\ell}\,m^{-\ell-\frac12}\,
\gamma\!\left(\ell+\tfrac12,\,m \chi_N^2\right).
\end{equation}
Substituting \eqref{eq:time_int_final} into \eqref{eq:avgS_gen_start} gives the general-$n$ average:
\begin{equation}\label{eq:avgS_gen_final}
\overline S^{(N)}=\ln 2-\frac{1}{2\lambda N\tau_{\text{R}}}
\sum_{m=1}^{\infty}\frac{1}{2m(2m-1)}
\sum_{\ell=0}^{2mn} c_{n,m,\ell}\; m^{-\ell-\frac12}\;
\gamma\!\left(\ell+\tfrac12,\, m(\lambda N\tau_{\text{R}})^2\right).
\end{equation}
For $n=0$, $L_0(x)=1$ so $c_{0,m,0}=1$ and $c_{0,m,\ell}=0$ for $\ell>0$.
Using $\gamma(1/2,z)=\sqrt{\pi}\,\erf\ (\sqrt{z})$,
\begin{equation}
\overline S^{(N)}=\ln 2-\frac{\sqrt{\pi}}{2\lambda N\tau_{\text{R}}}
\sum_{m=1}^{\infty}\frac{\erf\,\bigl(\chi_N\sqrt{m}\,\bigr)}{(2m)(2m-1)\sqrt{m}}.
\end{equation}

\twocolumngrid 

\bibliography{main}

\end{document}